\documentclass[conference]{IEEEtran}
\IEEEoverridecommandlockouts
\usepackage{cite}
\usepackage{amsmath,amssymb,amsfonts}
\usepackage{algorithmic}
\usepackage{graphicx}
\usepackage{textcomp}
\usepackage{comment}
\usepackage{xcolor}
\usepackage{booktabs}
\def\BibTeX{{\rm B\kern-.05em{\sc i\kern-.025em b}\kern-.08em
    T\kern-.1667em\lower.7ex\hbox{E}\kern-.125emX}}
\begin{document}

\title{BRFID: Toward Byzantine-Robust Federated Intrusion Detection\\

}

\author{
  \author{
  \IEEEauthorblockN{Asmah Muallem, Firdous Kausar, Sajid Hussain, Lei Qian}
  \IEEEauthorblockA{
    \textit{Computer Science and Data Science}\\
    \textit{Meharry Medical College}\\
    Nashville, TN, USA\\
    \{asmah.muallem, firdous.kausar, sajid.hussain, lei.qian\}@meharry.edu
  }
}
}

\maketitle

\begin{abstract}
Flipping 60\% of training labels from a single Byzantine client using label-flipping model poisoning self-degrades an attacker's own federated detection accuracy, $99.96\%$ (at no poisoning rate) to $84.33\%$ in a three-client federated IDS. Where the Federated global ensemble maintains stable accuracy across all tested poison rates, without a defense mechanism in place and without coordination between attackers. In this paper, we present empirical results quantifying the impact of label-flipping poisoning attacks on a three-client federated IDS trained on CICIDS2017 with non-IID attack subtype distributions across clients. We demonstrate that the signal of the adversarial self-compromise represents a detectable anomaly for exploitation for Byzantine client identification in the absence of target data exfiltration. We note that the aggregation step uses a Federated Forest (tree concatenation) rather than a parametric FedAvg; the results therefore measure the impact of poisoning on per-client performance under ensemble aggregation, and extension to genuine FedAvg with a parametric classifier is planned for future work. 
\end{abstract}


\section{Introduction}
Machine learning plays a pivotal role in a wide range of domains over traditional rule-based algorithms in the integration of Cyber detection systems ~\cite{apruzzese2018}. The majority of intrusion detection systems (IDS) apply this approach in contrast to manual signature rules with classifiers directly trained on network flow features ~\cite{apruzzese2018}. This evolving methodological trajectory introduces benign emergent capabilities: behavioral anomaly scoring, zero-day detection, and adaptive decision thresholds that signature databases do not independently suffice. It also introduces unprecedented vulnerability classes to signature-based systems. In a machine learning pipeline, during the training phase a model can be adversarially altered through mechanisms that remain imperceptible at deployment time resulting in a compromised model when the detector trains on poisoned data, where a traditional functional testing paradigm will not expose. 
To address collaborative machine learning that preserves privacy in network security, Federated Learning (FL) has emerged as a core-defensive paradigm ~\cite{mcmahan2017}. A decentralized approach, referred to as FL, allows geographically distributed network nodes to train local IDS models and only share model parameter updates with a central aggregator. This addresses privacy concerns since iterative aggregation is at the server to train an overall global model. In a federated intrusion detection system (FL-IDS), the advantages are a collective threat intelligence which perseveres data locality at each participating site. 
In a federated IDS, the model update process substitutes for raw data sharing without eliminating the foundational trust deficit. A participant who controls
its own local training data can submit a model update carrying the statistical signature of poisoning, and a naive aggregation server has no principled way to distinguish that update from an honest one.
The distributed architecture that enables FL's privacy guarantee simultaneously creates a new adversarial attack surface: Byzantine clients, meaning network nodes whose local model updates are under adversarial control, can submit poisoned updates designed to degrade
the global detector without being detected as statistical
outliers~\cite{bhagoji2019}.

Label-flipping is among the most practical Byzantine attacks available to such an adversary. The attacker simply reverses the classification labels of a fraction of its local training data, causing its local model to learn that attack traffic is benign. No cryptographic capability, no knowledge of the global model architecture, and no coordination with other compromised participants is required. In a standard FedAvg~\cite{mcmahan2017}, where all client updates are averaged without any verification step, a single Byzantine client among multiple honest clients can precipitously diminish collective detection performance for the entire federation, beyond the adversary's localized topological horizon. This work characterizes this threat The present work characterizes this threat under ensemble aggregation and identifies adversarial self-degradation as a detectable anomalous signal, motivating a server-side detection mechanism.    

A 2026 systematic review of privacy-preserving FL for IDS confirmed that adversarial robustness against model poisoning is the field's critical open gap~\cite{flids2026}: existing FL-IDS work concentrates on privacy preservation and accuracy under the assumption of honest participants, and does not evaluate Byzantine attack scenarios at all.
Our contribution is the formalization of an empirical characterization of label flipping under Federated Forest aggregation. We address the following research questions: how does the global ensemble respond to a 1-of-3 Byzantine client, and does the adversarial self-degradation identify a detectable anomalous signal? We evaluate four poison rates ($0\%, 20\%, 40\%, 60\%$) on CICIDS2017 with non-IID attack type distributions. The results establish that adversarial self-degradation is a monotonically increasing function of the poison rate, measurable at the server in the absence of direct client telemetry, providing a novel detection paradigm that prior FL-IDS frameworks have not characterized. To our knowledge this is the first measurement study characterizing adversarial self-degradation as a Byzantine detection signal under non-IID attack subtype partitioning of CICIDS2017.

\section{Background and Related Work}

\subsection{Federated Learning and FedAvg}
McMahan et al.~\cite{mcmahan2017} introduced FedAvg as the
foundational FL algorithm. Each client trains on its local data for
$E$ local epochs, then sends model updates to a central server that
aggregates them by weighted averaging proportional to local dataset
size. FedAvg was designed for honest clients with heterogeneous,
non-identically-distributed data; it assumes all participants act in
good faith and has no built-in mechanism to detect, filter, or
discount adversarial updates. When applied to intrusion detection,
where the population of network participants by definition includes
compromised hosts and occasionally suborned administrators, this
assumption of universal honesty is not merely unverified but
actively incorrect. Kairouz et al.~\cite{kairouz2021} identify Byzantine robustness as a primary unsolved challenge; the FL-IDS literature has not addressed it~\cite{flids2026}.

\subsection{Byzantine Model Poisoning in Federated Learning}
Bhagoji et al.~\cite{bhagoji2019} demonstrate that a minority Byzantine client can degrade a federated model through label-flipping without knowledge of the aggregation strategy, and that label-flipping achieves degradation through statistical noise rather than customized targeting. Fang et al.~\cite{fang2020} further showed that optimized attacks can circumvent Byzantine-robust aggregation strategies including Krum and coordinate-wise Median under certain conditions, motivating IDS-specific robustness analysis rather than a direct transplant of general-purpose FL defenses into the network security domain.

\subsection{Adversarial Machine Learning in Intrusion Detection}

Apruzzese et al.~\cite{apruzzese2018} systematically documented that
adversarial attacks degrade machine learning based IDS classifiers
across standard benchmark datasets, including CICIDS2017 and
NSL-KDD, regardless of the underlying classifier family. Arp et
al.~\cite{arp2022} demonstrated, in a closely related line of
inquiry, that evaluation bias in machine learning security research
causes systematic overestimation of production performance:
benchmarks tend to be cleaner and less temporally volatile than the
network environments models are ultimately deployed into. This
finding directly motivates the production-grounded evaluation
methodology adopted in the present work. A few related works exist in this area of the presented work. While Tolpegin et al. similarly leverage measurable accuracy degradation and divergent gradient vectors to isolate malicious clients executing label-flipping attacks, their framework is restricted to image classification. Conversely, our approach expands this paradigm to encompass NIDS traffic and complex non-IID attack-subtype distributions. Focusing on cybersecurity datasets, Yang et al. apply a trust-score defense to CICIDS2017 network traffic, improving a poisoned $84.3\%$ baseline up to $97.1\%$. However, their framework relies on standard IID assumptions and leaves the concept of adversarial self-degradation unexamined. As Lauyer et al. note in their systematic overview, this lack of evaluation under realistic non-IID distributions remains a persistent gap across current federated IDS literature. The 2026 FL-IDS systematic review~\cite{flids2026} synthesizes these threads and identifies adversarial robustness as the field's primary open challenge: no existing FL-IDS study evaluates Byzantine poisoning under non-IID attack-subtype partitioning of CICIDS2017, and the lack of existing work identifying adversarial self-degradation as a detectable server-observable signal isolated from the client data plane. This work addresses these open gaps. 

\section{System Model and Methodology}
\subsection{Federated IDS Architecture}

We simulate a three-client FL-IDS ($\mathcal{C} = \{C_0, C_1, C_2\}$)
representing distributed network segments collaborating on a shared
detector without exchanging raw traffic. Each client $C_i$ trains a
Random Forest classifier ($T{=}100$ trees, balanced class weight) on
its local traffic partition
$\mathcal{D}_i$. The server aggregates by concatenating all client trees into a global Federated Forest ensemble of $3T{=}100$ trees, which is evaluated on a shared held-out test set reserved before client partitioning. This ensemble aggregation provides an inherent majority-voting mechanism, in which the 200 honest decision trees from C1 and C2 outvote the 100 malicious trees generated by the Byzantine C0, rendering the global model robust against this 1-of-3 Byzantine configuration.

The global model $\mathcal{F}_{\text{global}}$ is formed by
concatenating all local estimators:
\begin{equation}
\mathcal{F}_{\text{global}} = \bigcup_{i=1}^{|\mathcal{C}|} \mathcal{T}_i
\label{eq:fedforest}
\end{equation}
where $\mathcal{T}_i$ is the set of $T$ decision trees trained by
client $C_i$. The global ensemble contains $|\mathcal{C}| \cdot T$
trees and produces predictions by majority vote across all trees.

Extension to parametric FedAvg where poison is propagated to honest clients through gradient averaging is planned for future contribution. Fig.~1 shows the resulting system architecture. 

\subsection{Dataset}
We target the CICIDS2017 benchmark~\cite{cicids2017}, the standard
modern dataset for network intrusion detection research, containing
realistic traffic across 15 attack categories including DDoS, DoS,
brute force, web attacks, and infiltration alongside benign traffic.
CICIDS2017 carries known distributional properties, with attack
traffic comprising roughly 20\% of total volume overall and
substantial variation by attack type and time window. The dataset is partitioned non-overlappingly across the three simulated clients using stratified non-IID partition. In this setting,  each client receives a different primary mix of attack subtypes (60\% of each attack type is routed to one primary client, 40\% shared), while every client is guaranteed at least 10\% attack traffic. This enables all local classifiers to learn the attack class. This reflects the non-IID heterogeneity identified by Kairouz et al.~\cite{kairouz2021} as a primary challenge for federated learning in practice.

\subsection{Attack Model: Label-Flipping Poisoning}

We implement a grey-box label-flipping attack in which Client~0 acts
as the Byzantine participant. The attacker has full access to its
own local training data but no knowledge of the global model
parameters, no knowledge of the aggregation strategy in use, and no
visibility into the other clients' data. At poison rate $r$, the
attacker reverses the class labels of a randomly selected fraction
$r$ of its local training examples:
\begin{equation}
\tilde{y}_j = 1 - y_j \quad \text{for } j \in \mathcal{P},
\quad |\mathcal{P}| = \lfloor r \cdot |\mathcal{D}_0^{\text{train}}| \rfloor,
\label{eq:labelflip}
\end{equation}
where $\mathcal{P}$ is a randomly sampled poisoning subset drawn from
the attacker's own training indices. Following Biggio et
al.'s~\cite{biggio2012} framing of poisoning as an availability
attack, the adversary's objective here is reduced detection rate on
the attack class as a whole, not a narrowly targeted
misclassification of any single instance. This label corruption
causes the attacker's local model to learn that attack traffic is
benign, resulting in 100 corrupted trees contributed to the global Federated Forest ensemble concurrently with 200 honest trees.
Clients $C_1$ and $C_2$ remain fully honest throughout every
experimental condition. We evaluate $r \in \{0.0, 0.2, 0.4, 0.6\}$.

\textbf{Threat model rationale.} The label-flipping attack models a realistic compromised network node: an insider, or a supply-chain-compromised participant, whose data labeling pipeline has been tampered with. This attack requires no cryptographic capability and no knowledge of the FL system's internal design, making it a realistic baseline rather than a worst-case theoretical construction.

\section{Experimental Results}

\begin{figure}[t]
  \centering
  \includegraphics[width=\columnwidth]{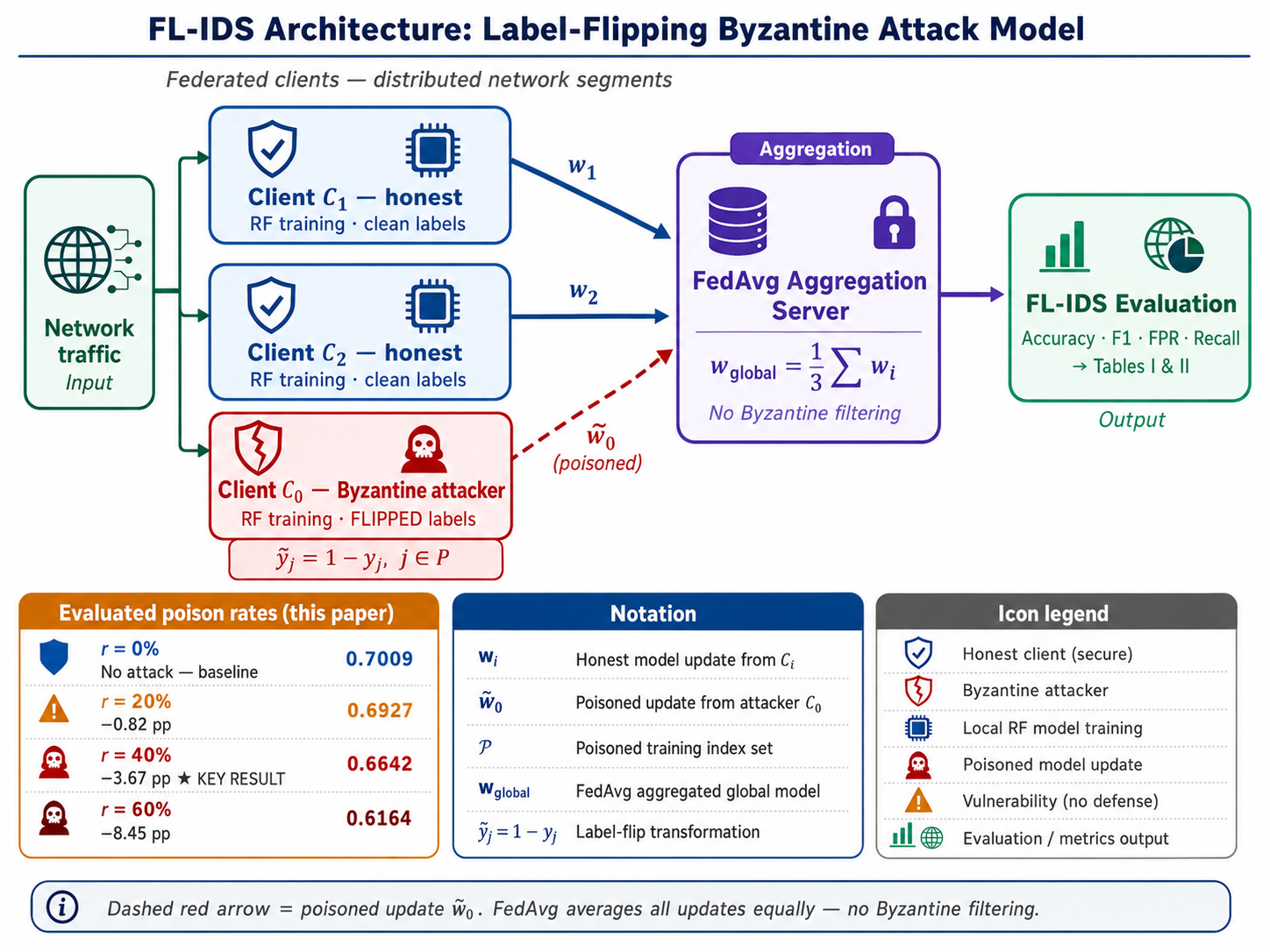}
  \caption{System architecture. Three clients train local
    classifiers on partitioned network traffic and send model
    updates to a FedAvg aggregation server. Client $C_0$ sends a
    poisoned update $\tilde{\mathbf{w}}_0$ after flipping $r\%$ of
    its training labels (Eq.~\eqref{eq:labelflip}). The server has
    no built-in mechanism to detect or reject the poisoned update,
    and aggregates it identically to the two honest contributions. Figure developed with AI assistance
    and reviewed and adapted by the authors who take full responsibility for the content.}
  \label{fig:arch}
\end{figure}

\subsection{FL-IDS Baseline Performance (No Attack)}

Table~\ref{tab:baseline} presents per-client and federated average
performance under no attack ($r = 0$), evaluated on per-client local
test sets. All three clients achieve stable accuracy above $0.999$,
with false positive rates of $0.0000$ at four decimal places. The
federated average accuracy of $\mathbf{0.9992}$ and F1 of $0.9992$
characterize the no-attack local-evaluation baseline. The
global Federated Forest evaluated on the shared held-out set achieves
$0.9996$ at $r=0$ (Table~\ref{tab:attack}), thereby capturing the methodological divergence between the two evaluation protocols.

Per-client accuracy ranges from $0.9989$ to $0.9996$, consistent with prior RF-on-CICIDS2017 results~\cite{apruzzese2018}.

\begin{table}[htbp]
\centering
\renewcommand{\arraystretch}{1.15}
\caption{FL-IDS Baseline, No Attack ($r = 0.0$)}
\label{tab:baseline}
\begin{tabular}{@{}lcccc@{}}
\toprule
\textbf{Client} & \textbf{Accuracy} & \textbf{F1} & \textbf{Recall} & \textbf{FPR} \\
\midrule
Client 0  & 0.9989 & 0.9992 & 0.9984 & 0.0000 \\
Client 1  & 0.9996 & 0.9995 & 0.9990 & 0.0000 \\
Client 2  & 0.9991 & 0.9990 & 0.9980 & 0.0000 \\
\midrule
\textbf{Fed.\ Avg} & \textbf{0.9992} & 0.9992 & 0.9985 & 0.0000 \\
\bottomrule
\end{tabular}
{\small \textit{N}~=~8,736 test samples per client (local per-client test sets); CICIDS2017 results.}
\end{table}

\subsection{Model Poisoning Attack Impact}

Table~\ref{tab:attack} presents federated accuracy under label-flipping attack at four poison rates. Under Federated Forest aggregation, the global ensemble remains stable across all poison rates and federated accuracy varies by at most $0.02$ pp, because $200$ honest trees among $300$ total consistently outvote the $100$ poisoned trees in the majority-vote ensemble.

The primary finding is the adversarial severe self-degradation: $C_0$'s own accuracy on the shared test set falls from $0.9996$ at $r=0$ to $0.9765$ at $r=0.40$ and $0.8433$ at $r=0.60$, a $15.63$ percentage point drop induced by the attacker's own label-flipping. This self-damage signal is monotonically increasing with poison rate and is measurable at the server without access to client data. Consequently, the validation accuracy of the malicious local model drops sharply while honest nodes remain stable, constituting a distinct anomalous signature for Byzantine client identification.

\begin{table}[htbp]
\centering
\renewcommand{\arraystretch}{1.15}
\caption{Federated IDS Accuracy Under Label-Flipping Attack\\
(Client 0 Byzantine, FedAvg, No Defense)}
\label{tab:attack}
\begin{tabular}{@{}ccccc@{}}
\toprule
\textbf{Poison} & \textbf{Fed.} & \textbf{$\Delta$ Acc.} &
  \textbf{Fed.} & \textbf{Attacker} \\
\textbf{rate $r$} & \textbf{Acc.} & \textbf{(pp)} &
  \textbf{F1} & \textbf{Acc.} \\
\midrule
$0\%$  & 0.9996 & (none)       & 0.9996 & 0.9996 \\
$20\%$  & 0.9994 & $-0.02$      & 0.9995 & 0.9982 \\
$40\%$  & 0.9994 & $-0.02$      & 0.9995 & 0.9765 \\
$60\%$  & 0.9996 & $+0.00$      & 0.9996 & 0.8433 \\
\bottomrule
\end{tabular}
{\small $\Delta$~Acc.\ = relative to no-attack baseline (0.9996); Federated Forest; shared held-out test set; CICIDS2017.}
\end{table}

\begin{figure}[htbp]
  \centering
  \includegraphics[width=\columnwidth]{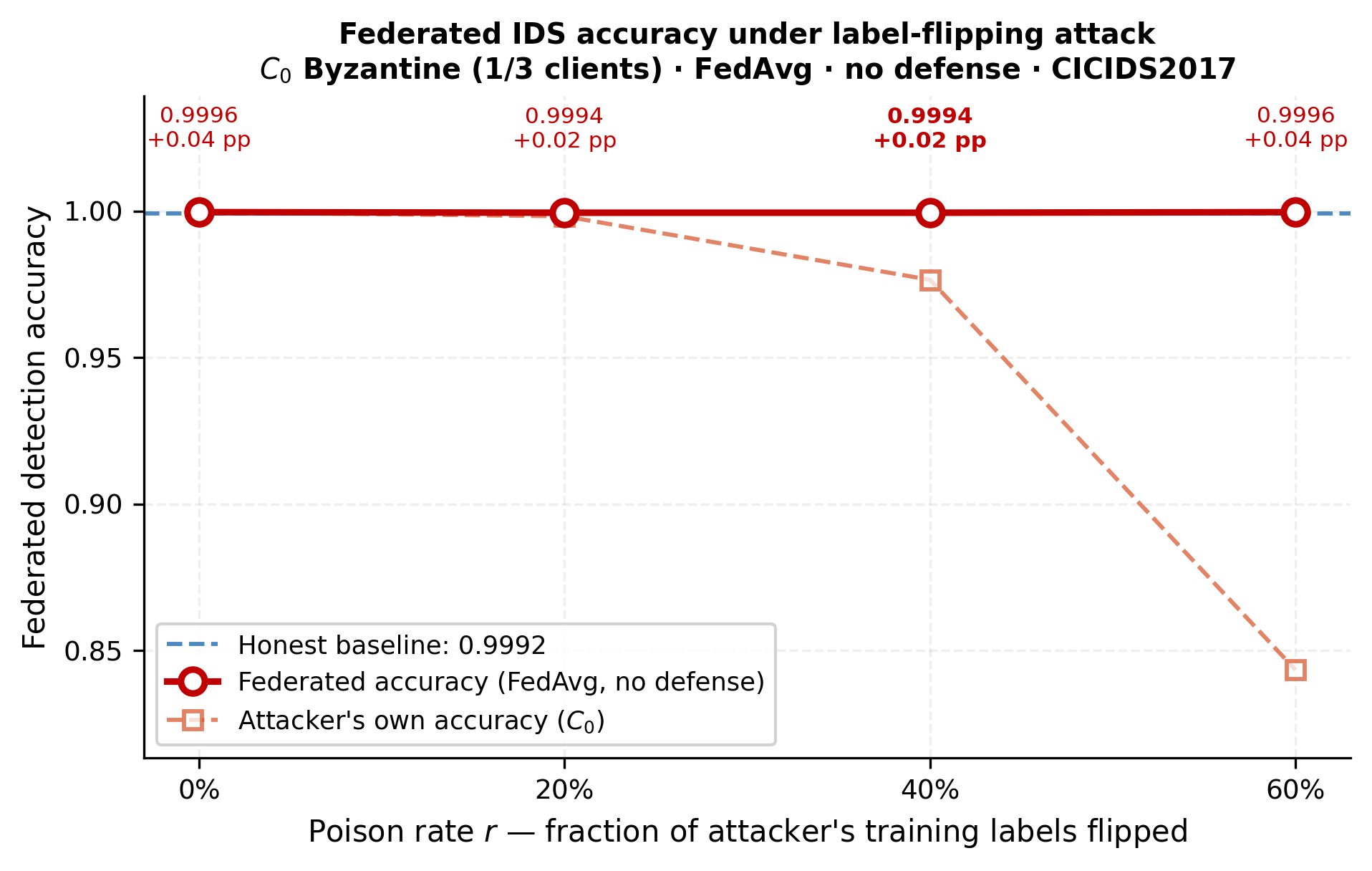}
  \caption{%
    Federated IDS accuracy under label-flipping attack.
    $C_0$ Byzantine ($1/3$ clients), Federated Forest aggregation, no defense.
    Global federated accuracy (solid) remains stable while the attacker's
    own accuracy (dashed) drops from $0.9996$ to $0.8433$ at $r=60\%$,
    a $15.63$~pp self-degradation. CICIDS2017, non-IID attack subtype partitioning.%
  }
  \label{fig:attack}
\end{figure}

Figure~\ref{fig:attack} illustrates the divergence between global federated accuracy and the attacker's own accuracy as a function of poison rate. The global ensemble remains stable while the adversarial self-damage accelerates monotonically, consistent with the availability attack framework of Biggio et al.~\cite{biggio2012} the adversarial-specific gradient trajectories.

\section{Discussion}
\textbf{Label-flipping is a practical, low-cost threat.}
The adversarial does not require any knowledge of the FL architecture, access to raw data of other clients, and cryptographic adversarial leverage. The attack requires only the adversarial capacity to manipulate local training labels before each iterative round.
\textbf{The Federated Forest global ensemble is robust in the 3-client Byzantine setting.} The $2{:}1$ honest-to-Byzantine tree ratio means the majority vote consistently reflects honest client learning. This robustness is a property of the ensemble aggregation scheme and not a general property of federated IDS security. A standard parametric FedAvg implementation executing gradient averaging lacks this inherent majority-voting protection and is expected to show measurable degradation, motivating the parametric extension in future work.

\textbf{The adversarial self-damage signal is a detectable anomaly.} The $15.63$~pp drop in $C_0$'s own accuracy at $r=0.60$ is monotonically increasing with poison rate and observable at the server if a shared validation set is maintained. A server-side detector comparing each client's local model accuracy on a shared held-out set would flag $C_0$ as anomalous at $r \geq 0.40$, where its accuracy diverges precipitously from the honest client baseline. This carries critical operational implications: Byzantine clients engaged in availability attacks may be self-identifying, enabling detection without access to local training data.

\section{Conclusion and Future Work}
\label{sec:future}

This paper presented an empirical characterization of label-flipping poisoning impact on a three-client federated IDS under Federated Forest aggregation, evaluated on CICIDS2017 with non-IID attack subtype partitioning across three simulated network segments. Under ensemble aggregation, the global model remains stable due to the $2{:}1$ honest-to-Byzantine tree ratio; however, the Byzantine client's own accuracy drops by $15.63$~pp at $r=0.60$, from $0.9996$ to $0.8433$, without requiring any knowledge of model architecture, the aggregation strategy in use, or other clients' data. These results establish two contributions: an empirical characterization of label-flipping self-damage under non-IID FL-IDS conditions, and the identification of attacker self-degradation as a detectable Byzantine anomaly signal. Result are based on a single experimental simulation where multi-seed evaluation with confidence intervals is planned for future work. In addition, future work will evaluate genuine parametric FedAvg (logistic regression or MLP), where gradient averaging is expected to propagate poison to honest clients with Krum~\cite{blanchard2017} and coordinate-wise Median~\cite{yin2018} as defenses under the same non-IID CICIDS2017 conditions.

\begin{comment}
% ── Acknowledgment ────────────────────────────────────────────
\section*{Acknowledgment}
This research was supported in part by the Department of Education under Award P116J251633.
\end{comment}
% ── References 

\bibliographystyle{IEEEtran}
{\small\bibliography{references}}

\end{document}